# Unifying Kibble–Zurek Mechanism in Weakly Driven Processes

**Pierre Nazé**

Instituto de Ciências Exatas e Naturais, Faculdade de Física, Universidade Federal do Pará, Av. Augusto Correa, 1, Guamá, Belém 66075-110, PA, Brazil; pierre.naze@icen.ufpa.br

**Abstract**

A description of the Kibble–Zurek mechanism with linear response theory has been done previously, but ad hoc hypotheses were used, such as the rate-dependent impulse window via the Zurek equation in the context of no driving in the relaxation time. In this work, I present a new framework where such hypotheses are unnecessary while preserving all the characteristics of the phenomenon. The Kibble-Zurek scaling obtained for the excess work is close to 2/5, a result that holds for open and thermally isolated systems with relaxation time that diverges at the critical point and the first zero of the relaxation function is finite. I exemplify the results using four different but significant types of scaling functions.

**Keywords:** Kibble-Zurek mechanism; linear response theory

## 1. Introduction

The Kibble–Zurek mechanism is a fundamental framework for understanding how systems driven through a critical point fall out of equilibrium, leading to the formation of defects and the emergence of universal scaling laws [1]. Initially developed in cosmology [2] and later applied to condensed matter physics [3], it has since been extended to a wide range of classical and quantum systems [4–16]. However, traditional formulations of the mechanism often rely on heuristic reasoning, for instance the introduction of a phenomenological freeze-out or impulse time, rather than a systematic derivation from microscopic dynamics [1].

This work develops a systematic approach to the Kibble–Zurek mechanism based entirely on linear response theory [17]. A central element of this approach is a new definition of the relaxation function derived directly from an appropriate relaxation time [17]. This construction naturally distinguishes between adiabatic and impulse contributions to the work performed during the driving, removing the need for phenomenological assumptions. The resulting framework reveals how scaling exponents emerge organically from equilibrium response properties and provides a principled derivation of the universal features of the mechanism. To demonstrate the framework, I study four representative models: the underdamped Brownian particle in a time-dependent harmonic trap, a system with a Bessel-type scaling function, the transverse-field quantum Ising chain, and a system with a sinc-type scaling function. For finite zeros of the relaxation function, the derived scaling laws of the excess work closely approximate a Kibble–Zurek exponent of 2/5.

This unified treatment provides a deeper theoretical foundation for the Kibble–Zurek mechanism and extends its applicability to a broader range of weakly driven systems. By explicitly linking equilibrium response functions to non-equilibrium thermodynamic behavior, this approach establishes a rigorous analytical path for studying critical dynamics in settings where numerical simulations or full dynamical treatments are challenging. Last but not least, compared to [18], the present work removes the need for the externally postulated impulse time or Zurek equation and instead derives the onset of impulsive behavior directly from the relaxation function.

## 2. Preliminaries

In this section, I define the definitions and notations to be used throughout this work.

### 2.1. Weakly Driven Systems

Consider a system with a time-dependent Hamiltonian $\mathcal{H}(\lambda(t))$, where $\lambda(t)$ is the external parameter that drives the system. Initially, the system is prepared at thermal equilibrium with a weakly coupled thermal reservoir with temperature $\beta^{-1}$. During the driving, the heat bath can be either with the system or not, being respectively called an open or thermally isolated system. I also consider that the external parameter is

$$\lambda(t) = \lambda_0 + g(t)\delta\lambda, \tag{1}$$

where $\lambda_0$ is its initial value, $\delta\lambda$ is its driving strength, and $g(t)$ is the protocol; in particular, the driving is weak, where I assume $|\delta\lambda g(t)/\lambda_0| \ll 1$. The process occurs during a switching time of the parameter $\tau$.

To measure the thermodynamic quantities, such as the thermodynamic work, internal energy, and absorbed heat at the end of the driving, one uses the relaxation function $\Psi_0(t)$ [17], given by

$$\Psi_0(t) = -\int \phi_0(t)dt + \mathcal{C}, \tag{2}$$

where $\phi_0$ is the response function [17] and the constant $\mathcal{C}$ can be chosen according to the characteristics of the system (being open or thermally isolated). For the first case, one can define the relaxation time

$$\tau_R = \int_0^\infty \frac{\Psi_0(t)}{\Psi_0(0)}dt. \tag{3}$$

For thermally isolated systems, such a definition does not work out, as the oscillatory behavior intrinsic to the relaxation function does not allow convergence in the above integral [19].

### 2.2. Kibble–Zurek Mechanism

The Kibble–Zurek mechanism provides a universal description of how a system driven across a continuous phase transition falls out of equilibrium due to the divergence of its relaxation time. When an external parameter $\alpha(t)$ is varied through its critical value $\alpha_0$, the relaxation time $\tau_R(\alpha(t))$ grows and eventually exceeds the characteristic time scale imposed by the driving. As a consequence, the system becomes unable to follow the protocol adiabatically and enters a regime commonly referred to as the impulse region, during which the state of the system remains effectively "frozen". This formation of an adiabatic–impulse–adiabatic sequence is responsible for the emergence of universal scaling laws in a broad class of classical and quantum systems.

In the traditional phenomenological formulation of the Kibble–Zurek mechanism, one assumes that the breakdown of adiabaticity occurs when the instantaneous relaxation time matches the inverse rate of change of the external parameter:

$$\tau_R(\alpha(\hat{t})) = \frac{\alpha(\hat{t})}{\dot{\alpha}(\hat{t})}, \tag{4}$$

which is often called the Zurek equation. This relation determines the crossing time $\hat{t}$ at which the dynamics switches from adiabatic to impulse behavior. The value of $\hat{t}$ controls the characteristic length and time scales of the resulting excitations and provides the foundation for scaling predictions such as the density of defects or excess work. Indeed, in this last case, the excess work scales as

$$\langle W\rangle \sim \tau^{-\gamma_{\mathrm{KZ}}}, \tag{5}$$

where $\gamma_{\mathrm{KZ}}$ is the Kibble–Zurek exponent. Typically, its value depends on the system; for instance, the transverse-field quantum Ising chain has $\gamma_{\mathrm{KZ}} = 1/2$ [1].

### *2.3. New Relaxation Time*

I will present a new definition of relaxation time which applies to both open and thermally isolated systems. It will be a fundamental piece in discussing incorporation of the Kibble–Zurek mechanism into weakly driven processes.

In [20], it was shown that new relaxation times emerge from new measurements of work via new relaxation functions in weakly driven processes. Our approach is to follow the reverse idea, that is, to find a new measure of work from the relaxation function defined by a new relaxation time, always in weakly driven processes. Then, consider the excess work of a thermally isolated or open system in such a regime [21]:

$$\langle W\rangle = \int_0^\tau \int_0^t \Psi_0(t-u)\dot{\lambda}(t/\tau)\dot{\lambda}(u/\tau)dudt. \tag{6}$$

For long switching times, $\tau \gg 1$, the driving can be put outside the integral [22]. Such an expression can be written in the following way:

$$\langle W\rangle = \Psi_0(0)\int_0^\tau \dot{\lambda}^2(t/\tau)\left[\int_0^t \frac{\Psi_0(u)}{\Psi_0(0)}du\right]dt \tag{7}$$

from where we can recognize the relaxation time

$$\tau_R(t) = \int_0^t \frac{\Psi_0(u)}{\Psi_0(0)}du \tag{8}$$

as the sum of all contributions of the relaxation function until time $t$. To work with only a number $\tau_R$, I choose the maximum one. This occurs at the first zero $t_0$, when the relaxation function vanishes. Therefore,

$$\tau_R = \int_0^{t_0} \frac{\Psi_0(t)}{\Psi_0(0)}dt. \tag{9}$$

To motivate this choice, consider the transverse-field quantum Ising chain, for which the Hamiltonian operator is

$$\mathcal{H} = -J\sum_{i=1}^N \sigma_i^x\sigma_{i+1}^x - \Gamma\sum_{i=1}^N \sigma_i^z. \tag{10}$$

Here, each of the $N$ spins has a vector $\vec{\sigma}_i := \sigma_i^x\mathbf{x} + \sigma_i^y\mathbf{y} + \sigma_i^z\mathbf{z}$ composed by the Pauli matrices. The parameter $J$ is the coupling energy and $\Gamma$ is the transverse magnetic field. In addition, the system is subjected to periodic boundary conditions and to an even number of spins, and has an initial temperature $T = 0$. The critical point occurs at $\Gamma_0 = J$, in the thermodynamic limit, $N \to \infty$. The relaxation function is given by [18]

$$\Psi_0(N,t) = \sum_{n=1}^{N/2} \frac{16J^2}{\epsilon_n^3}\sin\left(\frac{(2n-1)}{N}\pi\right)^2\cos\left(\frac{2\epsilon_n}{\hbar}t\right), \tag{11}$$

with

$$\epsilon_n = 2\sqrt{J^2+\Gamma_0^2-2\Gamma_0 J\cos\left(\frac{(2n-1)}{N}\pi\right)}. \tag{12}$$

The divergence behavior for $\Psi_0(N,0)$ was already discussed in [18]. In the critical point $\Gamma_0 = J$, $\epsilon_n$ becomes

$$\epsilon_n = 2J\sqrt{2 - 2\cos\left(\frac{(2n-1)}{N}\pi\right)} \tag{13}$$

in its thermodynamic limit

$$\epsilon(k) = 2J\sqrt{2 - 2\cos(k)}, \tag{14}$$

which closes the first gap for $k \to 0^+$. Using Equation (9), the relaxation time is

$$\tau_R(k) = \frac{\hbar}{2\epsilon(k)}, \tag{15}$$

which is the relaxation time normally used in the Kibble–Zurek mechanism of the transverse-field quantum Ising chain [3]. Thus, this example provides a concrete illustration of why the first zero of the relaxation function is the appropriate characteristic time scale. In the thermodynamic limit, the relaxation function is a weighted superposition of cosine modes with frequency $2\epsilon(k)/\hbar$, where the gap $\epsilon(k)$ vanishes as $k \to 0^+$ at criticality. The first zero of this relaxation function occurs at $t_0 = \pi\hbar/2\epsilon(k)$, which precisely matches the relaxation time $\tau_R = \hbar/2\epsilon(k)$ widely used in traditional Kibble–Zurek analyses.

This identification is not incidental; it reflects the fact that the breakdown of adiabaticity in the Ising chain is controlled by the slowest mode. The coherence if this mode is lost when its fundamental oscillation completes one quarter of a cycle. Thus, the use of the first zero is not merely convenient but is directly tied to the microscopic dynamics that control defect formation and Kibble–Zurek mechanism scaling in this model. This provides a strong microscopic validation for extending the same definition to general systems. Another motivation for this choice, perhaps stronger, will become clear when I discuss the new work generated by this relaxation time.

Now, the new relaxation time can be rewritten as

$$\overline{\tau}_R = \int_0^\infty \frac{\overline{\Psi}_0(t)}{\overline{\Psi}_0(0)}dt, \tag{16}$$

where

$$\overline{\Psi}_0(t) = \Psi_0(t)\theta(t_0 - t)\theta(t_0 + t), \tag{17}$$

in which $\theta(t)$ is the Heaviside theta. Note that the new relaxation function is even; in addition, it will provide the new work associated with the new relaxation time $\overline{\tau}_R$ of the system.

### 2.4. New Work

Let us now deduce the new work $\langle\overline{W}\rangle$ associated with this new relaxation function. Observing that

$$\overline{\Psi}_0(t-u) = \Psi_0(t-u)\theta(t_0 - t)\theta(t_0 - u)\theta(t_0 + t)\theta(t_0 + u) \tag{18}$$

and applying

$$\langle\overline{W}\rangle = \frac{1}{2}\int_0^\tau\int_0^\tau \overline{\Psi}_0(t-u)\dot{\lambda}(t/\tau)\dot{\lambda}(u/\tau)dudt, \tag{19}$$

we have

$$\langle\overline{W}\rangle = \begin{cases} \frac{1}{2}\int_0^\tau\int_0^\tau \Psi_0(t-u)\dot{\lambda}(t/\tau)\dot{\lambda}(u/\tau)dudt, & t_0 \geq \tau \\ \frac{1}{2}\int_0^{t_0}\int_0^{t_0} \Psi_0(t-u)\dot{\lambda}(t/\tau)\dot{\lambda}(u/\tau)dudt, & t_0 \leq \tau \end{cases}. \tag{20}$$

For $t_0 \leq \tau$ and $t_0 \geq \tau$, I call this new work adiabatic and impulse work, respectively. Two situations can happen here: $t_0$ is either infinite or finite. In the first case, the situation is identical to the usual definition of the excess work [21]. However, in the second case, if the switching time is large compared to the first zero, then the new work is restricted to a time window inside the whole process duration. This is quite similar to what happens in the Kibble–Zurek phenomenological description [1]. In particular, the choice of the first zero becomes clear now; only positive parts of the relaxation function contribute to the impulse work, avoiding possible cancellations of negative parts in that amount. In other words, the impulse work becomes the largest possible. Indeed, the first zero corresponds to the first time at which the response ceases to be constructive.

It is important to emphasize that linear response theory does not impose a unique definition of the relaxation time. Many characteristic times exist, including integrated times, spectral linewidths, and inverse decay rates; each highlights different physical aspects of relaxation. The definition adopted here, based on the integral up to the first zero of the relaxation function, is chosen because it captures the maximal interval during which the response remains constructive and directly comparable to the standard adiabatic–impulse separation of the Kibble–Zurek mechanism. In this sense, the definition is not presented as the only possible relaxation time but rather as the one in which the operational meaning aligns naturally with the physical onset of non-adiabaticity and in which the properties reproduce known Kibble–Zurek mechanism properties without additional assumptions. By explicitly acknowledging this choice as physically motivated rather than axiomatically imposed, the approach remains transparent and broadly applicable while retaining internal consistency. Last but not least, observe that different definitions of relaxation time across various types of systems may exhibit distinct features, such as in the scalings of observables related to the switching time.

In the following sections, I start to discuss the driven processes associated with the crossing of the critical point in the Kibble–Zurek mechanism.

## 3. Quench Processes

For open and thermally isolated systems, let us observe the behavior of quench processes, that is, processes in which the values of the initial conditions $\alpha$ are in the critical point $\alpha_0$, producing divergence of the relaxation time

$$\lim_{\alpha\to\alpha_0} \tau_R(\alpha) = \infty. \tag{21}$$

Observe that one can translate the expression of the relaxation time into the following universal expression:

$$\int_0^{t_0(\alpha)/\tau_R(\alpha)} \frac{\Psi_0(\alpha,\tau_R t)}{\Psi_0(\alpha,0)} dt = 1. \tag{22}$$

If the relaxation time diverges in this limit, then in order for such an equation to be consistent the first zero needs to be infinity as well; therefore,

$$t_0(\alpha_0) = \tau_R(\alpha_0) = \infty. \tag{23}$$

Also, at this limit $\alpha \to \alpha_0$, the equation becomes

$$\Psi_0(\alpha_0,\infty) = \Psi_0(\alpha_0,0). \tag{24}$$

In other words, the integral of the response function is

$$\int_0^\infty \phi_0(\alpha_0,t)dt = 0. \tag{25}$$

We have one of the two cases: either $\phi_0(\alpha_0, t) = 0$ and $\Psi_0(\alpha_0, t)$ is a constant, or both functions decay oscillating. However, since the relaxation function has the same value at the beginning and end, being an even function where $\Psi_0(0) \neq 0$, then only the first case is true. Thus, the relaxation function is constant. This fact can be understood as the frozen state present in the Kibble–Zurek mechanism phenomenology, where the system no longer responds and does not relax as well.

Observe now that the relaxation function does not depend on $t$, which implies that the excess work does not nullify in the quasistatic process unless it is equal to zero. This is out of scope, since the system would not present any driving action. To guarantee that the Second Law of Thermodynamics still works, that is,

$$\lim_{\tau\to\infty}(W_{\mathrm{qs}} + W_{\mathrm{ex}}) = W_{\mathrm{qs}}, \tag{26}$$

the quasistatic work needs to diverge as well. This should happen at every order. In the first order, it is automatically satisfied. In particular, in the second order one has

$$\lim_{\tau\to\infty}\left(-\frac{\Psi_0(\alpha_0,0)}{2} + W_{\mathrm{ex}}^{(2)}\right) = -\frac{\Psi_0(\alpha_0,0)}{2}. \tag{27}$$

Therefore,

$$\Psi_0(\alpha_0,t) = \Psi_0(\alpha_0,0) = \Psi_0(\alpha_0,\infty) = \infty. \tag{28}$$

In particular, the new work will diverge as well:

$$\langle \overline{W} \rangle = \infty. \tag{29}$$

As shown in [18] for the transverse-field quantum Ising chain, linear response theory can be used to describe situations close to the Kibble–Zurek mechanism (small number of spins) when the strength of the perturbation is sufficiently small. However, inside the critical point, the situation changes from water to wine. Considering that the same effect of the strength of the perturbation of the transverse-ified quantum Ising chain remains here, the extreme results presented above demonstrate that linear response theory breaks down quantitatively, since the driving strength must be equal to zero. This is the case even though the interpretation of the results, like the existence of a frozen state (as observed in the Kibble–Zurek mechanism phenomenology), remains valid; in other words, linear response theory predicts and breaks down in the frozen state.

It is quite odd to think of a frozen state observing the behavior only at the critical point. Indeed, the Kibble–Zurek mechanism states that the frozen state begins at the onset of the impulse case. However, in the context of weakly driven processes, the relaxation time does not change with time, depending only on the initial parameter values. Therefore, the frozen state must be at the initial values where the criticality of the relaxation time occurs. This justifies why we choose to observe the frozen state at the critical point. Additionally, it is natural to think that a near-divergence in the relaxation time will persist close to the critical point.

## 4. Close to the Critical Point

I now analyze the situation near the critical point, which is the usual case in which the Kibble–Zurek mechanism and weakly driven processes coincide. As we saw in the previous section, because the relaxation function diverges at the critical point, the hypotheses used next will be fully justified.

### 4.1. Scaling Functions

Close to the critical point, I suppose that the relaxation function is composed of a term of magnitude with a critical exponent and a scaling function. Therefore,

$$\Psi_0(\alpha,t) \approx \Psi_0(\alpha,0) f\left(\frac{t}{\tau_R(\alpha)}\right), \tag{30}$$

where

$$f(t) = \lim_{\alpha\to\alpha_0} \frac{\Psi_0(\alpha,\tau_R(\alpha)t)}{\Psi_0(\alpha,0)} \tag{31}$$

with $f(0) = 1$. The value of $f(\infty)$ depends on whether the system is open or thermally isolated. The term $\Psi_0(\alpha,0)$ carries the term with the critical exponent, while $f(t)$ is the scaling function. Observe that in the limit $\alpha \to \alpha_0$, the scaling function loses its parameter dependence, while it can maintain its time dependence; indeed, it depends only on the macroscopic characteristics of the relaxation time. In addition, because it comes from the relaxation function, it is a finite, even, positive-definite, and analytical function. It is expected that the term $\Psi_0(\alpha,0)$ starts to diverge close to the critical point due to the existence of a critical exponent.

The response function is approximately

$$\phi_0(\alpha,t) \approx -\frac{\Psi_0(\alpha,0)}{\tau_R(\alpha)} f'\left(\frac{t}{\tau_R(\alpha)}\right), \tag{32}$$

with $f'(0) = 0$. Observe that in the critical point, $\phi_0(\alpha_0,t) = 0$, which implies

$$\lim_{\alpha\to\alpha_0} \frac{\Psi_0(\alpha,0)}{\tau_R(\alpha)} f'\left(\frac{t}{\tau_R(\alpha)}\right) = 0, \tag{33}$$

showing a particularity between the critical exponent and the way in which the scaling function decorrelates. Observe that our definition of the scaling function should preserve the relaxation time

$$\frac{\int_0^{\bar{t}_0(\alpha)/\tau_R(\alpha)} f(t)dt}{\int_0^{t_0(\alpha)} f(t)dt} = 1, \tag{34}$$

where $\bar{t}_0$ is the critical zero of $f(t/\tau_R)$ and $t_0$ is the critical zero of $f(t)$. Such a relation implies

$$\tau_R(\alpha) = \frac{\bar{t}_0(\alpha)}{t_0(\alpha)}. \tag{35}$$

Note that in the description of the scaling function, the microscopic aspect of the relaxation time is unnecessary; only its divergence is important. Therefore, in the case of open systems, some possible examples of scaling functions are

$$f(t) = \exp(-|t|), \quad f(t) = J_0(t), \tag{36}$$

with $\bar{t}_0/\tau_R = \infty$ for the first one and $\bar{t}_0/\tau_R = j_{0,1}$ for the second, which is the first zero of $J_0(t)$. Here, $J_0(t)$ is the Bessel function of the first kind of order 0. For thermally isolated systems, a possible example is

$$f(t) = \cos(t), \tag{37}$$

with $\bar{t}_0/\tau_R = \pi/2$. This will be the scaling function of the transverse-field quantum Ising chain (see Section 7.2.1).

*4.2. Expansions of the Scaling Function*

Let us study some expansions of the approximated relaxation function related to the relaxation time. I am going to focus on the scaling function. For $t/\tau_R(\alpha) \ll 1$, one has

$$f\left(\frac{t}{\tau_R(\alpha)}\right) \sim 1. \tag{38}$$

Considering now $t/\tau_R(\alpha) \gg 1$, I use the asymptotics of the scaling function for large $t/\tau_R(\alpha)$:

$$f\left(\frac{t}{\tau_R(\alpha)}\right) \sim \lim_{t/\tau_R(\alpha)\gg 1} f\left(\frac{t}{\tau_R(\alpha)}\right). \tag{39}$$

For instance, for the case of open systems [21] this will be

$$f\left(\frac{t}{\tau_R(\alpha)}\right) \sim \frac{\tau_R(\alpha)}{t}, \tag{40}$$

while thermally isolated and gapped systems [19] it will be

$$f\left(\frac{t}{\tau_R(\alpha)}\right) \sim \left(\frac{\tau_R(\alpha)}{t}\right)^2. \tag{41}$$

## 5. Scalings at the New Work

To start, consider the linear driving

$$\alpha(t) = \alpha_0 + \frac{t}{\tau}\delta\alpha \tag{42}$$

with $t \in [-\tau/2, \tau/2]$, where the initial condition is $\alpha_i = \alpha_0 - \delta\alpha/2$ with $\delta\alpha/\alpha_0 \ll 1$. Let us now observe the scalings of the new work in the switching time $\tau$.

*5.1. Case $\tau/\bar{t}_0 \geq 1$*

When $\tau/\bar{t}_0 \geq 1$, we are in the adiabatic case. We have

$$\langle \overline{W} \rangle = \frac{\delta\alpha^2}{2\tau^2} \int_{-\bar{t}_0}^{\bar{t}_0} \int_{-\bar{t}_0}^{\bar{t}_0} \Psi_0(\alpha_i, t-u)dudt. \tag{43}$$

In addition,

$$\langle \overline{W} \rangle \sim \frac{\bar{t}_0^2}{\tau^2} \int_{-1}^{1} \int_{-1}^{1} f(t_0(t-u))dudt. \tag{44}$$

The integral will be a number independent of $\tau$. Therefore, the new rate scaling under such conditions will be

$$\langle \overline{W} \rangle \sim \left(\frac{\tau_R}{\tau}\right)^2, \tag{45}$$

where I have used $\bar{t}_0 \propto \tau_R$.

*5.2. Case $\tau/\bar{t}_0 \lesssim 1$*

When $\tau/\bar{t}_0 \lesssim 1$, we just enter the impulse case. This situation is where the Kibble–Zurek mechanism starts to happen, with an appropriate scaling for the excess work [1]. One has

$$\langle \overline{W} \rangle = \frac{\delta\alpha^2}{2\tau^2} \int_{-\tau/2}^{\tau/2} \int_{-\tau/2}^{\tau/2} \Psi_0(\alpha_i, t-u)dudt. \tag{46}$$

Additionally,

$$\langle \overline{W} \rangle \sim \int_{-1/2}^{1/2} \int_{-1/2}^{1/2} f\left(\frac{\tau}{\tau_R}(t-u)\right)dudt. \tag{47}$$

We will consider the following supposition:

$$\langle\overline{W}\rangle\left(\frac{\tau}{\tau_R}\right)=A\left(\frac{\tau_R}{\tau}\right)^{\gamma_{\rm KZ}} \tag{48}$$

where $A$ is a constant that will be appropriately chosen later and $\gamma_{\rm KZ}$ is the Kibble–Zurek exponent. Rewriting as

$$\langle\overline{W}\rangle\left(\frac{t_0\tau}{\bar{t}_0}\right)=A\left(\frac{\bar{t}_0}{t_0\tau}\right)^{\gamma_{\rm KZ}} \tag{49}$$

and considering an expansion around $\tau=\bar{t}_0$, it is possible to show under the appropriate choice of $A$ that the exponent is

$$\gamma_{\rm KZ}=-\frac{\langle\overline{W}\rangle'(t_0)}{\langle\overline{W}\rangle(t_0)}t_0. \tag{50}$$

This is the Kibble–Zurek exponent, which holds for any system that presents a divergent relaxation time defined by Equation (9) at the critical point. This approach furnishes a new dynamical conceptualization of the Kibble–Zurek exponent. It measures the ratio of the first-order approximation of the excess work at the first zero with respect to its actual value at the same point. Therefore, by observing its value, one can infer how fast the excess work is changing at the first zero, in other words, how fast the adiabaticity has been lost. Note that the Kibble–Zurek exponent does not depend on the strength of the perturbation of the process. In addition, it encodes information about the equilibrium state, since it depends on the first zero of the scaling function, which is a correlation calculated at equilibrium. Observe that the equality $\tau=\bar{t}_0=t_0\tau_R(\alpha_i)$ is nothing more than the Zurek equation, which expresses the rate at which the system starts to become impulsive. This observation shows that our approach is consistent and recovers the Kibble–Zurek mechanism phenomenology.

It is also possible to show, under the approximation of the relaxation function until second order, that it furnishes the rational number

$$\gamma_{\rm KZ}\approx\frac{2}{5}. \tag{51}$$

Relying on such a second-order approximation is comprehensible, since we want to analyze the instant at which the system enters the impulse regime, that is, when the new work exhibits nonlinear behavior. Last but not least, it is essential to recognize that the Kibble–Zurek scaling is a local one, and improves its extension with higher relaxation time; indeed, in the second-order approximation of the scaling function, the second derivative of the logarithm of the new work is $\sim 1/\tau_R^2$, indicating that the derivative changes very slowly. However, this matter is very subtle, since the increasing of the relaxation time can turn linear response inadequate to describe such phenomena quantitatively. This will be better explained in Section 6.

### *5.3. Case $\tau/t_0\leq 1$ and $\tau/\tau_R\ll 1$*

For $\tau/t_0\leq 1$, one has

$$\langle\overline{W}\rangle\sim\frac{\delta\alpha^2}{\tau^2}\int_{-\tau/2}^{\tau/2}\int_{-\tau/2}^{t}\Psi_0(\alpha_i,t-u)dudt. \tag{52}$$

Additionally,

$$\langle\overline{W}\rangle\sim\int_{-1/2}^{1/2}\int_{-1/2}^{1/2}f\left(\frac{\tau}{\tau_R}(t-u)\right)dudt. \tag{53}$$

Because $\tau/\tau_R$ is very close to zero and the integral limits are close to one, I use the approximation (38). The new rate scaling of the new work under such conditions will be

$$\langle\overline{W}\rangle \sim \tau^0, \tag{54}$$

which is independent of $\tau_R$. In summary, the new scalings will be

$$\langle\overline{W}\rangle \sim \begin{cases} \tau^0, & \tau/\bar{t}_0 \leq 1, \tau/\tau_R \ll 1 \\ \tau^{-\gamma_{KZ}}, & \tau/\bar{t}_0 \lesssim 1 \\ \tau^{-2}, & \tau/\bar{t}_0 \geq 1. \end{cases} \tag{55}$$

## 6. Range of Validity in Linear Response Theory

The derivation of the presented Kibble–Zurek scaling relies entirely on linear response theory; therefore, its domain of applicability deserves clarification. The assumptions required for the use of linear response theory impose intrinsic restrictions on how close to the critical point the system may be driven as well as on the form of the resulting scaling exponents.

First, the driving strength must remain small throughout the protocol, in the sense that

$$\frac{\delta\alpha}{\alpha_0} \ll 1, \tag{56}$$

so that higher-order terms in the response expansion remain negligible. Although this condition is standard in weakly driven processes, its meaning is more subtle near the critical point; the relevant scale is not only the magnitude of the parameter change but whether the perturbation stays small compared to the characteristic distance from criticality at which adiabaticity breaks down. In other words, the perturbation must remain weak relative to the time-dependent gap of the relaxation dynamics.

Second, as shown explicitly in Section 3, linear response theory ceases to be quantitatively valid at the critical point. In this limit, both the relaxation time and the magnitude of the relaxation function diverge, while the response function vanishes identically. The resulting frozen state is fully consistent with Kibble–Zurek phenomenology, but the theory predicts a divergent quasistatic work unless the driving amplitude tends to zero. This establishes that linear response theory cannot describe dynamics inside the critical point, but remains reliable for processes that traverse the critical region with a finite (though possibly large) relaxation time that does not change in time with the driving.

Third, the Kibble–Zurek exponent obtained here should be interpreted as a local scaling exponent valid near the onset of the impulse regime. Because the new work inherits the second-order truncation of the relaxation function, the exponent is determined by the curvature of the scaling function around its first zero. As a consequence, the value $2/5$ emerges generically whenever $t_0 < \infty$, whereas recovering a familiar exponent of the full Kibble–Zurek mechanism for a typical system requires going beyond linear response. Indeed, the fact that the Kibble–Zurek exponent does not depend on the strength of the driving reveals its limitations.

Finally, because the second derivative of $\ln\langle\overline{W}\rangle$ behaves as $1/\tau_R^2$, the local scaling improves as the relaxation time increases, confirming that the present approach becomes increasingly accurate for slow relaxation but finite $\tau_R$. In summary, the linear response description developed here applies to weakly driven processes sufficiently close to but not inside the critical point, with a very small perturbation as the relaxation time becomes larger but finite. Within this regime, the theory captures the emergence of a Kibble–Zurek scaling of the excess work at the onset of the impulse region.

## 7. Examples

### *7.1. Open Systems*

I start by presenting examples of open systems.

#### 7.1.1. Underdamped Brownian Motion

Consider an underdamped Brownian motion driven by the stiffening parameter according to the Langevin equation [23]

$$m\ddot{x}(t) + \gamma\dot{x}(t) + \omega_0^2(t)x(t) = \eta(t), \tag{57}$$

where $m$ is the mass of the particle, $\gamma$ is the friction parameter, $\omega_0(t)$ is the time-dependent natural frequency, and $x(t)$ is the position of the particle. The random force $\eta(t)$ is white noise such that

$$\langle\eta(t)\rangle = 0, \quad \langle\eta(t)\eta(t')\rangle = \frac{2m\gamma}{\beta}\delta(t-t'), \tag{58}$$

where $\beta$ is proportional to the inverse initial temperature $T$. Using linear response theory [17], the relaxation function is given by

$$\frac{\Psi_0(t)}{\Psi_0(0)} = \frac{e^{-\gamma|t|}}{\omega^2}\left[2\omega_0^2 + \left(\omega^2 - 2\omega_0^2\right)\cos\omega t + \gamma\omega\sin\omega|t|\right], \tag{59}$$

$$\Psi_0(0) = \frac{1}{2m^2\beta\omega_0^4}, \tag{60}$$

where $\omega = \sqrt{4\omega_0^2 - \gamma^2}$ is a positive number. The relaxation time is

$$\tau_R(\gamma,\omega_0) = \frac{\gamma^2 + \omega_0^2}{2\gamma\omega_0^2}. \tag{61}$$

We are going to see that the conditions to achieve Kibble–Zurek mechanism are the limits $\gamma \to 0^+$ and $\omega_0 \to 0^+$, at the same rates, with $\gamma/\omega_0 = 1 < 2$. Observe that such conditions imply $\omega \geq 0$.

Considering initial situations where the limits $\gamma \to 0^+$ and $\omega_0 \to 0^+$, at the same rates and with $\gamma/\omega_0 = 1$, one has

$$\lim_{\substack{\gamma,\omega_0\to 0^+\\ \gamma/\omega_0=1}} \tau_R(\gamma,\omega_0) = +\infty, \tag{62}$$

$$\lim_{\substack{\gamma,\omega_0\to 0^+\\ \gamma/\omega_0=1}} \Psi_0(0) = +\infty, \tag{63}$$

$$\lim_{\substack{\gamma,\omega_0\to 0^+\\ \gamma/\omega_0=1}} \frac{\Psi_0(t)}{\Psi_0(0)} = 1. \tag{64}$$

Such an example illustrates the same effects of the Kibble–Zurek mechanism; the relaxation time and the norm of the relaxation function diverge and the relaxation function is a constant, exhibiting a frozen state of the system in which the system takes a very long time to equilibrate after the driving. The scaling function is

$$\lim_{\substack{\gamma,\omega_0\to 0^+\\ \gamma/\omega_0=1}} \frac{\Psi_0(\tau_R t)}{\Psi_0(0)} = e^{-|t|}. \tag{65}$$

From now on, for large $\tau_R$, the relaxation function used is

$$\Psi_0(t) \sim e^{-\frac{|t|}{\tau_R}}, \tag{66}$$

for which $\bar{t}_0 = \infty$. Such an infinite first zero will make all the difference; indeed, for $\tau_R = 100$, the new work has the following scaling in the switching times:

$$\langle \overline{W} \rangle \sim \begin{cases} \tau^0, & \tau/\bar{t}_0 \leq 1, \tau/\tau_R \ll 1 \\ \tau^{-\gamma_{\rm KZ}}, & \tau/\bar{t}_0 \lesssim 1 \end{cases} \tag{67}$$

where

$$\gamma_{\rm KZ} = 1, \tag{68}$$

according to Equation (50). Figure 1 corroborates the result. The scaling always seen in open systems with non-negative relaxation functions [21] is Kibble–Zurek mechanism scaling in the context of a diverging relaxation time. Indeed, from the mathematical point of view such divergence is always possible, allowing the existence of a Kibble–Zurek mechanism. Further work will be developed to understand the microscopic aspects of such a divergence in underdamped Brownian motion.

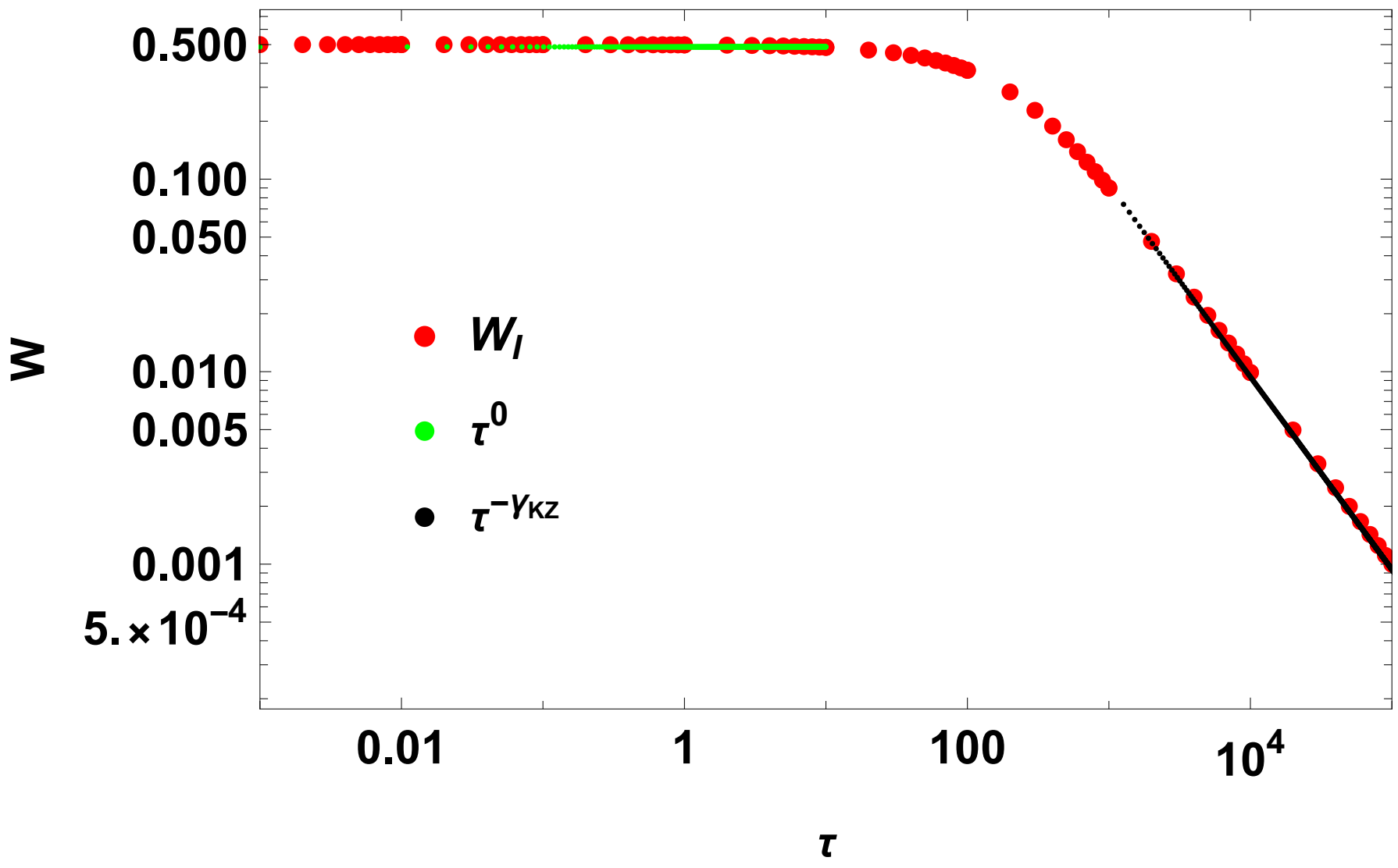

**Figure 1.** New work and new scalings for $e^{-|t|}$. The Kibble–Zurek coefficient is $\gamma_{\rm KZ} = 1$, using $\tau_R = 100$.

### 7.1.2. Bessel Function $J_0(t/\tau_R)$

Let us suppose that our system is a spin model subject to a heat bath and obeying Glauber dynamics [24]. Suppose also that it presents a Bessel function as a scaling function

$$\Psi_0(t) \sim J_0\left(\frac{t}{\tau_R}\right), \tag{69}$$

with $\bar{t}_0 = j_{0,1}\tau_R$. I additionally assume that such a system presents an intrinsic mechanism of divergence in its relaxation time. Assuming $\tau_R = 100$, the impulse work has the following scaling in the switching times:

$$\langle \overline{W} \rangle \sim \begin{cases} \tau^0, & \tau/\bar{t}_0 \leq 1, \tau/\tau_R \ll 1 \\ \tau^{-\gamma_{\rm KZ}}, & \tau/\bar{t}_0 \lesssim 1 \\ \tau^{-2}, & \tau/\bar{t}_0 \geq 1 \end{cases} \tag{70}$$

where the exponent $\nu$ is not possible to achieve due to $j_{0,1}$ being a small number. The Kibble–Zurek exponent is

$$\gamma_{\mathrm{KZ}} = 1 - \frac{2J_1(j_{0,1})}{j_{0,1}F[\{\frac{1}{2}\},\{\frac{3}{2},2\},-\frac{j_{0,1}^2}{4}]} \approx 0.454, \tag{71}$$

where $F$ is the generalized hypergeometric function. Figure 2 illustrates the result. As previewed, $\gamma_{\mathrm{KZ}} \approx 2/5$ for finite first zeros of the scaling function.

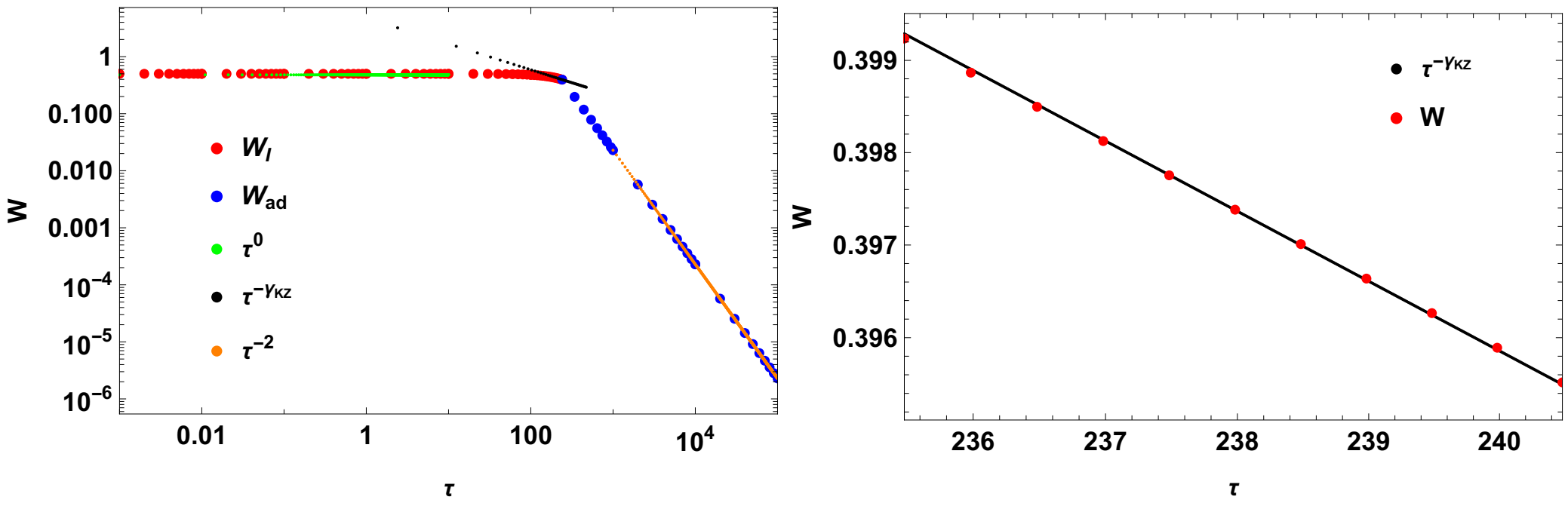

**Figure 2.** New work and new scalings for $J_0(t)$. The Kibble–Zurek coefficient is $\gamma_{\mathrm{KZ}} \approx 0.454$, using $\tau_R = 100$.

### *7.2. Thermally Isolated Systems*

Now, I present examples of thermally isolated systems.

#### 7.2.1. Transverse-Field Quantum Ising Chain

For more details about the transverse-field quantum Ising chain, see Section 1. Its scaling function is

$$f(t) = \lim_{N\to\infty} \frac{\Psi_0(N,\tau_R(k)t)}{\Psi_0(N,0)} = \cos(t). \tag{72}$$

Assuming $\tau_R = 100$, the impulse work has the following scaling in the switching times

$$\langle \overline{W} \rangle \sim \begin{cases} \tau^0, & \tau/\bar{t}_0 \leq 1, \tau/\tau_R \ll 1 \\ \tau^{-\gamma_{\mathrm{KZ}}}, & \tau/\bar{t}_0 \lesssim 1 \\ \tau^{-2}, & \tau/\bar{t}_0 \geq 1 \end{cases} \tag{73}$$

where

$$\gamma_{\mathrm{KZ}} = 2 - \frac{\pi}{2} \approx 0.429. \tag{74}$$

Figure 3 corroborates the result. Observe that a comparison with the classical Kibble–Zurek exponent $1/2$ is out of scope, since the system is not at the conditions of the Kibble–Zurek mechanism ($\tau_R = 100$). The Kibble–Zurek exponents should differ, as indeed they did. Again, for a finite first zero of the scaling function, the result corroborates $\gamma_{\mathrm{KZ}} \approx 2/5$.

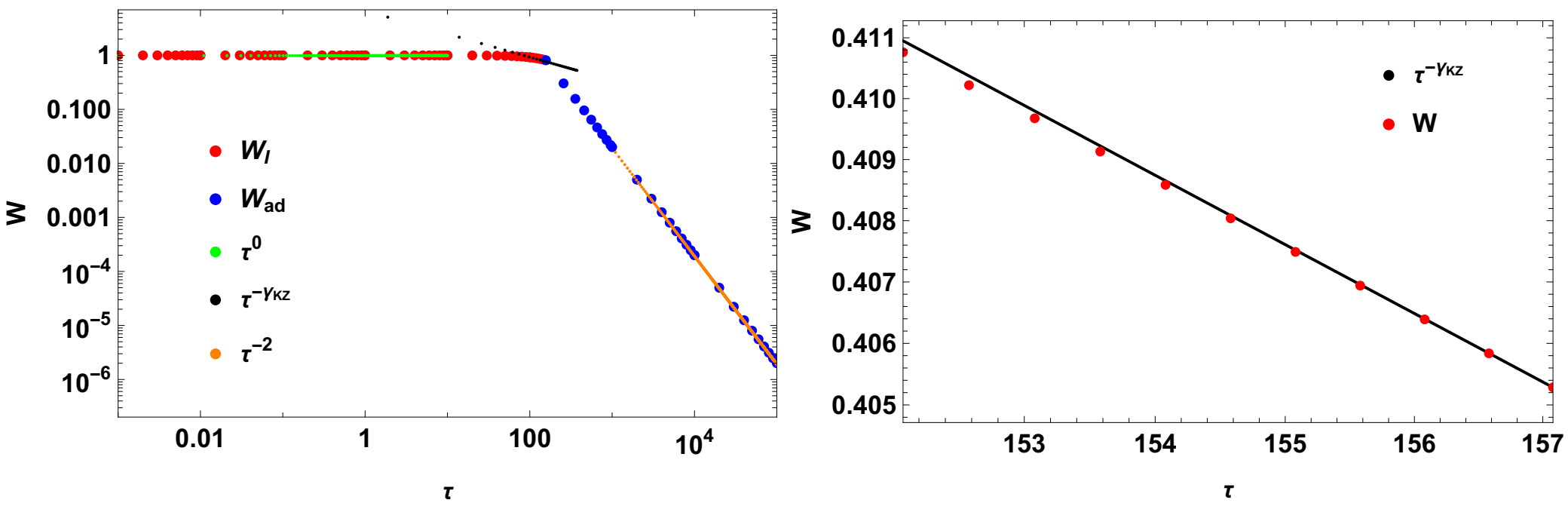

**Figure 3.** New work and new scalings for the cos $(t)$. The Kibble–Zurek coefficient is $\gamma_{\rm KZ} \approx 0.429$, using $\tau_R = 100$.

7.2.2. Function sinc$(t)$

Now, suppose a scaling function of the following form:

$$f(t) = \frac{\sin(t)}{t} \tag{75}$$

which is a kind of cosine relaxation function system that behaves as an open one when using the time-averaged work [20]. Again, I assume that such a system contains an intrinsic divergence mechanism for its relaxation time. Assuming $\tau_R = 100$, the impulse work has the following scaling in the switching times:

$$\langle\overline{W}\rangle \sim \begin{cases} \tau^0, & \tau/\bar{t}_0 \leq 1, \tau/\tau_R \ll 1 \\ \tau^{-\gamma_{\rm KZ}}, & \tau/\bar{t}_0 \lesssim 1 \\ \tau^{-2}, & \tau/\bar{t}_0 \geq 1 \end{cases} \tag{76}$$

where

$$\gamma_{\rm KZ} = \frac{\pi s(\pi)}{\pi s(\pi) - 2} - \frac{4}{\pi s(\pi) - 2} \approx 0.476, \tag{77}$$

with

$$s(x) = \int_0^x \frac{\sin(t)}{t} dt. \tag{78}$$

Figure 4 illustrates the result. The result corroborates $\gamma_{\rm KZ} \approx 2/5$.

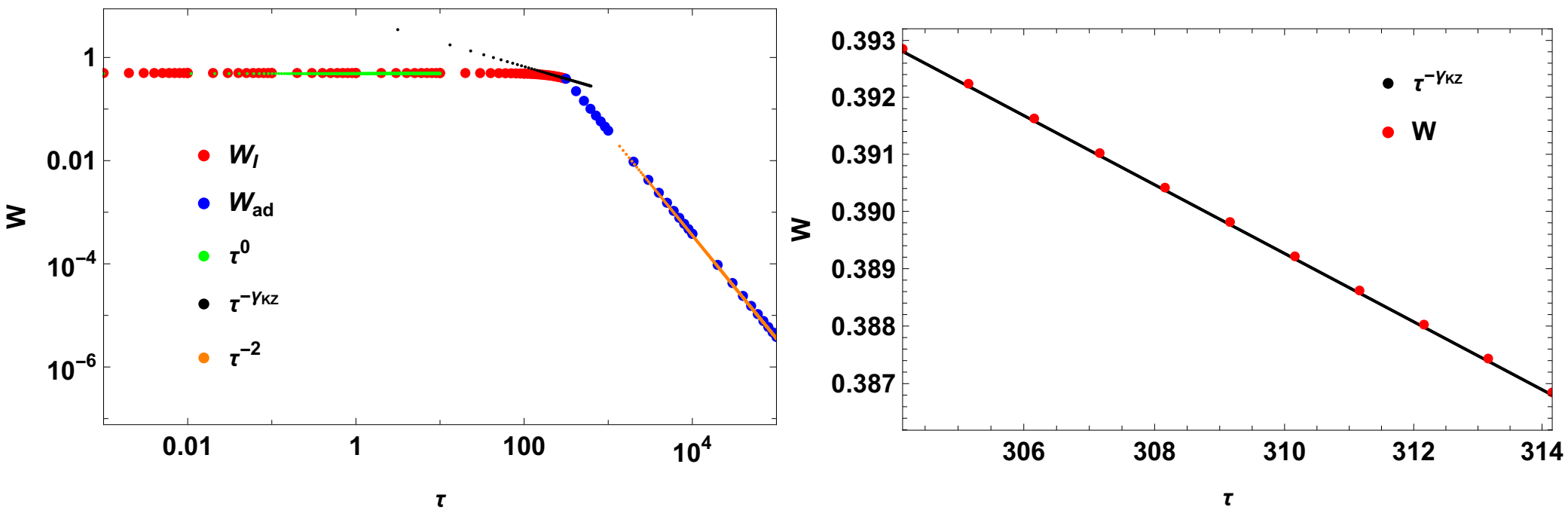

**Figure 4.** New work and scaling for the sinc$(t)$. The Kibble–Zurek coefficient is $\gamma_{\rm KZ} \approx 0.476$, using $\tau_R = 100$.

## 8. Final Remarks

In this work, I have developed a linear response framework capable of describing the nonequilibrium scaling that emerges when a system is driven across a continuous phase transition. The key idea was to identify the first zero $t_0$ of the equilibrium relaxation function as the natural boundary separating adiabatic and impulse behavior. This choice provides a microscopic and model-independent criterion for the breakdown of adiabaticity, replacing the phenomenological construction traditionally used in the Kibble–Zurek mechanism. The analysis leads to three main results.

First, I introduce a definition of relaxation time that applies to both open and isolated systems and that remains finite even when the relaxation function oscillates, thereby avoiding divergences that arise in conventional definitions. This construction recovers the standard relaxation timescale in models such as the transverse-field quantum Ising chain, providing a direct bridge between equilibrium correlations and critical dynamics.

Second, I show that the new work generated during a weak linear protocol exhibits a crossover controlled by the dimensionless ratio $\tau/t_0$. When the switching time becomes comparable to $t_0$, the system enters a regime in which the new work scales as a power law, in close analogy with the Kibble–Zurek scenario (indeed, this is analogous to the Zurek equation). This demonstrates that the adiabatic–impulse structure is already encoded in the equilibrium relaxation function and does not require any external phenomenology.

Third, expanding the relaxation function around its first zero allows the extraction of an effective Kibble–Zurek exponent for the excess work. For all relaxation functions possessing a finite first zero, I find a value close to $2/5$ within the domain of linear response. This value reflects the second-order structure of the relaxation function near $t_0$ and represents a universal prediction of linear response. In contrast, monotonic relaxations with no finite zero lead to the exponent equaling 1. Importantly, recovering the full Kibble–Zurek exponent requires going beyond linear response, confirming that the difference arises from the truncation inherent in the linear response framework.

I have also clarified the range of validity of the theory. Linear response provides a quantitatively reliable description as long as the driving remains sufficiently weak for a relaxation time large but finite; at the critical point, the relaxation function becomes time-independent and the quasistatic work diverges, signaling the expected breakdown of linear response theory. Nevertheless, the qualitative features of the frozen state agree with the standard Kibble–Zurek picture.

Overall, this work shows that the essential ingredients of Kibble–Zurek physics, such as adiabatic breakdown, impulse behavior, and scaling of excess work, can be derived directly from equilibrium two-time correlations without invoking non-equilibrium phenomenology. The framework developed here can be extended to nonlinear protocols, higher-order response theory, and systems with long-range memory kernels or colored noise, offering a promising route toward a fully microscopic formulation of universal critical dynamics.

Finally, the advantage of dealing with scaling functions is the independence of the microscopic aspects that contribute to the quench processes, maintaining only their dependence on the divergence of the relaxation time. Many systems can have the same scaling function, and consequently the same exponent, in universal classes. Despite this disparity between the scaling functions, the Kibble–Zurek scaling for the excess work seems to follow the exponents shown below, according to the critical zero of the relaxation function defined by Equation (9):

$$\gamma_{\text{KZ}} \approx \begin{cases} 1, & t_0 = \infty \\ \frac{2}{5}, & t_0 < \infty \end{cases} \tag{79}$$

which appears to be the only criterion to observe in determining the Kibble–Zurek exponent for the excess work. It is worth noting that the Kibble–Zurek exponents depend on how the relaxation time is defined and the observables studied; different definitions or observables may lead to different exponents.

**Funding:** This research received no external funding.

**Data Availability Statement:** The data presented in this study are openly available in https://github.com/pnaze/KZWD, accessed on 5 January 2026.

**Acknowledgments:** I thank Sebastian Deffner for enlightening discussions.

**Conflicts of Interest:** The author declares no conflicts of interest.